\documentclass[aps,prl,twocolumn,groupedaddress]{revtex4}
\usepackage{ulem}
\usepackage[english]{babel}  
\usepackage{graphicx}   

\usepackage{verbatim}
\newcommand{\chalmersMC}{$^{1}$ Department of Microtechnology and Nanoscience, MC2, Chalmers University of Technology,  SE-41296 G\"{o}teborg, Sweden}
\advance\textheight by 1 cm
\begin{document}
\title{Temperature stability of intersubband transitions in AlN/GaN quantum wells}
\author{Kristian Berland}
\affiliation{\chalmersMC}
\author{Martin Stattin}
\affiliation{\chalmersMC}
\author{Rashid Farivar}
\affiliation{\chalmersMC}
\author{D. M. S. Sultan}
\affiliation{\chalmersMC}
\author{Per~Hyldgaard}
\affiliation{\chalmersMC}
\author{Anders Larsson}
\affiliation{\chalmersMC}
\author{Shu Min Wang}
\affiliation{\chalmersMC}
\author{Thorvald G. Andersson}
\affiliation{\chalmersMC}


\date{\today}

\begin{abstract}
Temperature dependence of intersubband transitions in AlN/GaN multiple quantum wells grown with molecular beam epitaxy
is investigated both by absorption studies at different temperatures and modeling of conduction-band electrons. For the absorption study, the sample is heated in increments up to $400^\circ$C. The self-consistent Schr\"odinger-Poisson modeling includes temperature effects of the band-gap and the influence of thermal expansion on the piezoelectric field. We find that the intersubband absorption energy decreases only by $\sim 6$~meV at $400^\circ$C relative to its room temperature value. 
\end{abstract}


\maketitle

Intersubband transitions in nitride-based semiconductor heterostructures hold great promise for optoelectronic devices. The high band-offset between gallium nitride (GaN) and aluminum nitride (AlN) enables intersubband transitions in the near-infrared regime ($\lambda= 1$-$4 \,{\rm \mu m}$). Thus, intersubband devices such as modulators, detectors, and quantum cascade lasers (QCLs) have the potential to operate at wavelengths useful for fiberoptic communication. QCLs also have the potential to be used when measuring characteristic absorption of small molecules. There are several studies of intersubband absorption in AlN/GaN heterostructures.\cite{Gmachl2000, Kishino2002, Helman2003,aggerstam:2007,  AbsSaphire, kandaswamy:ISB} These typically show a peak at 500 to 900 meV with full width half maximum (FWHM) in the range of 60 to 200 meV. An intersubband device should operate at and above 300 K, often with the condition of a negligible change in transition energy. Temperature dependence of the transition energy, up to room temperature, has been reported for {\it intersubband} absorption in InAs/AlSb multiple quantum well (MQW) structures \cite{larrabee:3936} and for {\it interband} photoluminescence in nitride-based AlN/GaN MQW structures.\cite{Lin2004} The huge piezoelectric fields introduce an additional temperature-dependent effect in the nitride MQW structures, because the different thermal expansions of AlN and GaN induce a temperature dependent strain in the structures. 

In this letter, we demonstrate that the peak position of intersubband transitions red shifts only $\sim 6\,\mathrm{meV}$ as the temperature of the sample is increased from $25^\circ$ to $400^\circ{\rm C}$. This aspect of thermal robustness could be vital for the operation of nitride-based QCLs since their ultrafast LO-scattering rates and high intersubband energy would cause them to heat significantly. The temperature effects are studied for MQW structures both experimentally by measuring the absorption at different temperatures, and theoretically by a self-consistent solution to the Schr\"odinger-Poisson equations for the conduction-band electrons. The nitride-semiconductor heterostructures are characterized by huge internal electric fields, which arise from the exceptionally large spontaneous and piezoelectric fields. A proper account of these fields are crucial for an accurate description and characterization of the quantum well states, and they require that the complete structure is considered in the design of heterostructures.

The set of MQW structures (A - C) are grown in a Varian Gen II modular molecular beam epitaxy (MBE) system. Elemental gallium and aluminum are evaporated from thermal effusion cells, and nitrogen from a plasma source. For substrates, we use  2.5 $\rm \mu m$ thick silicon-doped GaN templates on $2^{''}$ sapphire substrates. 
We grow crack-free MQW structures on top of a silicon-doped GaN buffer layer at $760 \pm 5^\circ{\rm C}$ and thereafter cap the structure with a 500 nm silicon doped GaN layer. \cite{Brandt2001,Ive2005}

For the absorption study, the MQW structures are cut into $5\times7$ mm large chips and polished into a
standard multi-pass geometry with $45^\circ$ beveled input and exit edges. The temperature is set within $\pm 5$ $^\circ$C using a resistive heater and measured with a PT100 probe.
The transmitted light is focused onto a TE-cooled InGaAs detector (1.2-2.75 $\mu$m, ThorLabs
PDA10DT). The absorption in the MQW structures is measured using a Bruker IFS 55 FTIR Spectrometer.
A (ThorLabs LPNIR100) linear (0.65-2 $\mu$m, 30 dB ER) polarizer is used to select
transverse-magnetic (TM) polarized light.
We measure the transmission at room temperature and in 50 $^\circ$C increments between 50 and
400 $^\circ$C. Two measurement runs are made: one where the temperature is increased between each increment,  and one where it is decreased. For background reference at each temperature, a sapphire dummy sample is used. 

The intersubband absorption energies and conduction band profile are calculated using the Schr\"odinger-Poisson equation for envelope-functions within the effective-mass approximation:\cite{Gunna2007}
\begin{equation}
\left[ -{ \hbar^2 \over 2 } {\partial \over \partial z } \frac{1}{m(z,E)} {\partial \over \partial z } + V[n](z) \right] \psi_i(z) = E_i \psi_i(z)
\end{equation}
The conduction band profile, $V[n](z)$, takes into account the electrostatic potential arising from the interface charges, the electrostatic interactions between electrons (Hartree-term), and exchange-correlation in the local-density approximation.\cite{AbsSaphire,LDA:Bloss,Hedin1971} For the intersubband absorption energies, we include depolarization shift and excitonic effects. \cite{AbsSaphire,Gunna2007} For physical parameters of the nitrides, we use those recently recommended by Vurgaftman {\it et al}.\cite{Vurgaftman2007} 
Temperature effects enter the calculations through their influence on AlN and GaN band gaps, the thermal expansion influencing the piezoelectric fields, and the two-dimensional Fermi distribution of the electrons. 

The Varshni relation accounts for the change in band gap $E_{\rm gap}(T) = E_0 - \alpha T^2/(T+\beta)\,$ with temperature $T$. The parameters of the two nitrides differ significantly: for GaN $\alpha_{\rm GaN}=0.914\, \rm{mev K^{-1}}\,,  \beta_{\rm GaN}=825\,{\rm K}$; while for AlN $\alpha_{\rm AlN}=2.63\, \rm{mev K^{-1}}\,,  \beta_{\rm AlN}=2082\, {\rm K}$.  We keep the relative conduction/valence band offset fixed at all temperatures in the calculation. We further assume that the MQW structure is strained to the lattice parameter of the GaN substrate. The effective conduction band offset is therefore set to $\Delta E_c\approx 2\, {\rm eV} $ at room temperature.\cite{AbsSaphire,Bernardini1998} 

The polarization field within a layer is given by the sum of the spontaneous and piezoelectric polarization:
\begin{equation}
P_z(T)= P_z^{\rm SP} + 2 \frac{a(T)-a_0(T)}{a_0(T)} \left( e_{31} - e_{33} \frac{c_{13}}{c_{33}} \right). \label{eq:pol}
\end{equation}
Here $c_{13}$ and $c_{33}$ are elastic coefficients, $e_{31}$ $e_{33}$ are piezoelectric coefficients, and $a(T)$ is the layer lattice constant, while $a_0(T)$ is the lattice constant of the corresponding bulk material. Since the lattice constants of the active part of the MQW is set to that of the GaN substrate, the wells are unstrained, and the piezoelectric field contributes to the polarization fields only within the barriers. 
For the thermal expansion, we use the relation specified in Ref.~\onlinecite{Iwanaga2000}.

The Schr\"odinger equation is solved using the finite-difference method.\cite{FDM} This scheme benefits from an expansion of the non-parabolicity correction. We use an open-source routine to solve the resulting eigenvalue problem.\cite{scipy}  Potential mixing are used to dampen charge oscillations in the self-consistent cycle. 

\begin{figure}[t]
\centering
\includegraphics[width=8.5cm]{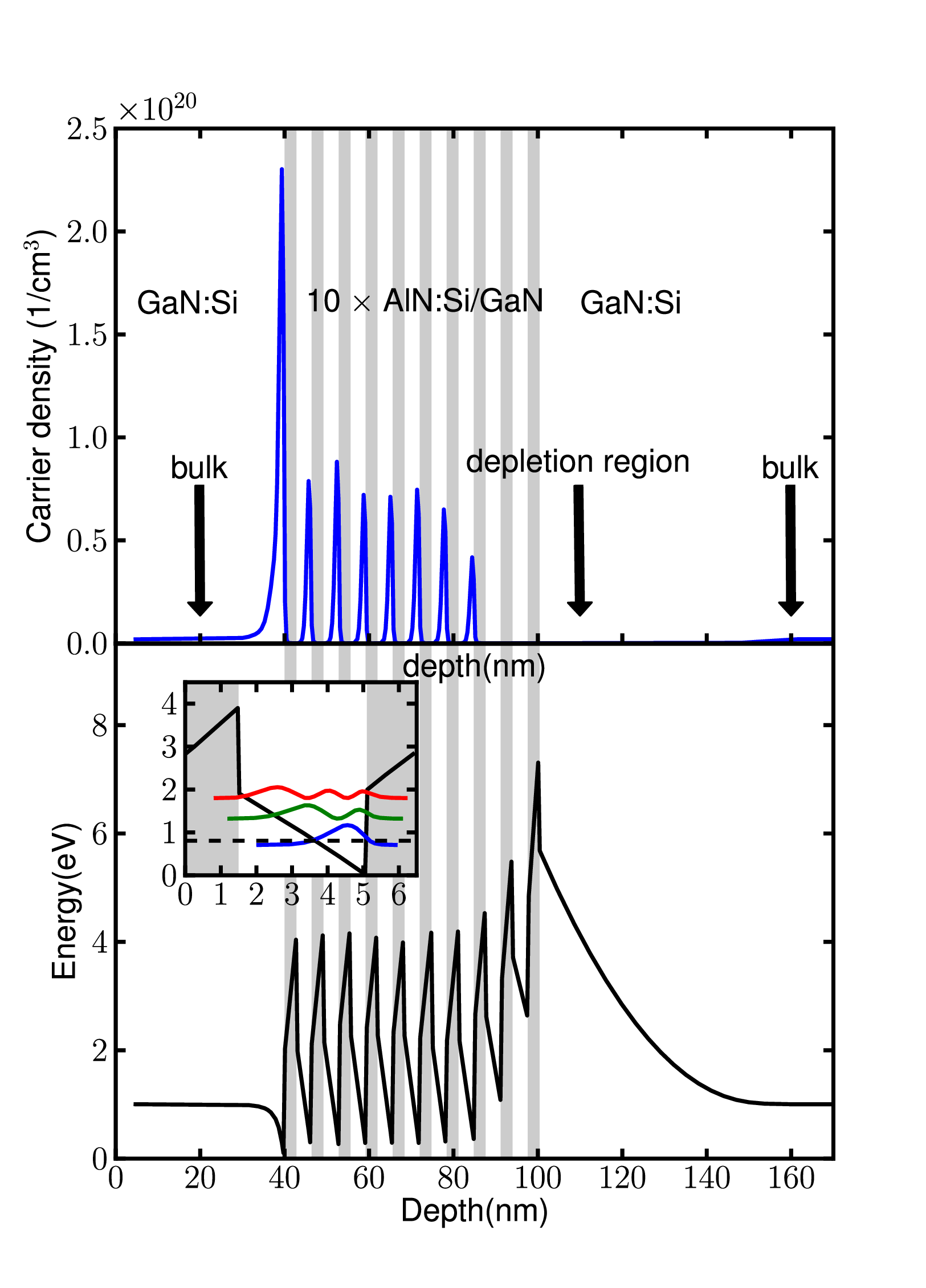}
\caption{The carrier density (upper panel) and conduction band profile (lower profile) of a ten period AlN/GaN multiple quantum well structure (sample B) with doped AlN barriers (gray background)  and doped GaN cladding layers generated by the Schr\"odinger-Poisson solver. The two rightmost wells are completely depleted and do not contribute to the absorption. The inset illustrates the three lowest bound state of a period of the MQW structure. The dashed line gives the room-temperature Fermi level.}
\label{fig:Profile}
\end{figure}

\begin{table}[h]
\label{Tab:ExpCalc}
\begin{ruledtabular}
\caption{Characteristics of the samples A, B, and C at room temperature: barrier and well thickness, experimental and calculated absorption peak positions.}
\begin{tabular}{llll}
Sample & barrier/{\bf well}   &measured peak & calculated peak\\
&thickness ({\rm nm}) &   postion $({\rm meV})$ &  position $({\rm meV})$  \\
\hline
A  &   4.2/{\bf 3.3}   &704  &  690   \\
B &   2.8/{\bf 3.6}   & 626 &  612   \\
C  &   3.1/{\bf 2.4}  & 739 &   712  \\
\end{tabular}
\end{ruledtabular}
\end{table}
Table I specifies the thickness of the AlN barriers and GaN wells for the three samples. It also gives the experimental and calculated (for a single period) room-temperature peak position of the absorption curve.
The barriers are delta-doped with a sheet-donor concentration of $n_d=7\cdot 10^{12}~\rm{cm}^{-2}$, while the GaN cladding layers are doped to $n_d=2\cdot~10^{18}~\rm{cm}^{-3}$.

When examining or designing heterostructures in materials with huge internal fields, it is essential to consider the full structure, as the long-range electric fields could  distort, or even deplete the active part of the structure.\cite{Gunna2007} To investigate these matters for the MQW structure of this study, we solve the Schr\"odinger-Poisson equation for the full structure in equlibrium (with a common Fermi level) and make sure to use a unit cell size large enough to get bulk-like behavior far from the central cell. 

Figure \ref{fig:Profile} shows the calculated carrier density and conduction band profile for the multiple-quantum well (MQW) structure of sample B. The light gray background indicates the position of the doped AlN barriers. At the first GaN/AlN interface (from the left), a two-dimensional electron gas is formed. Following the right AlN/GaN interface, there is a long electrostatic tail, depleted of conduction electrons. Moreover, the two rightmost wells are completely depleted, while the third from the right is partially depleted. Between these two regions, the potential profile of the MQW structure approximately repeats itself and this region should therefore dominate the absorption spectra. Thus, we can focus our calculations on a single period of the structure.
The inset of Fig. \ref{fig:Profile} displays the potential profile and the three energetically lowest states for a single period of the MQW structure. The calculated and experimental intersubband absorption energies are given in Table.~I. 

\begin{figure}[b]
\centering
  \includegraphics[width=8.5cm]{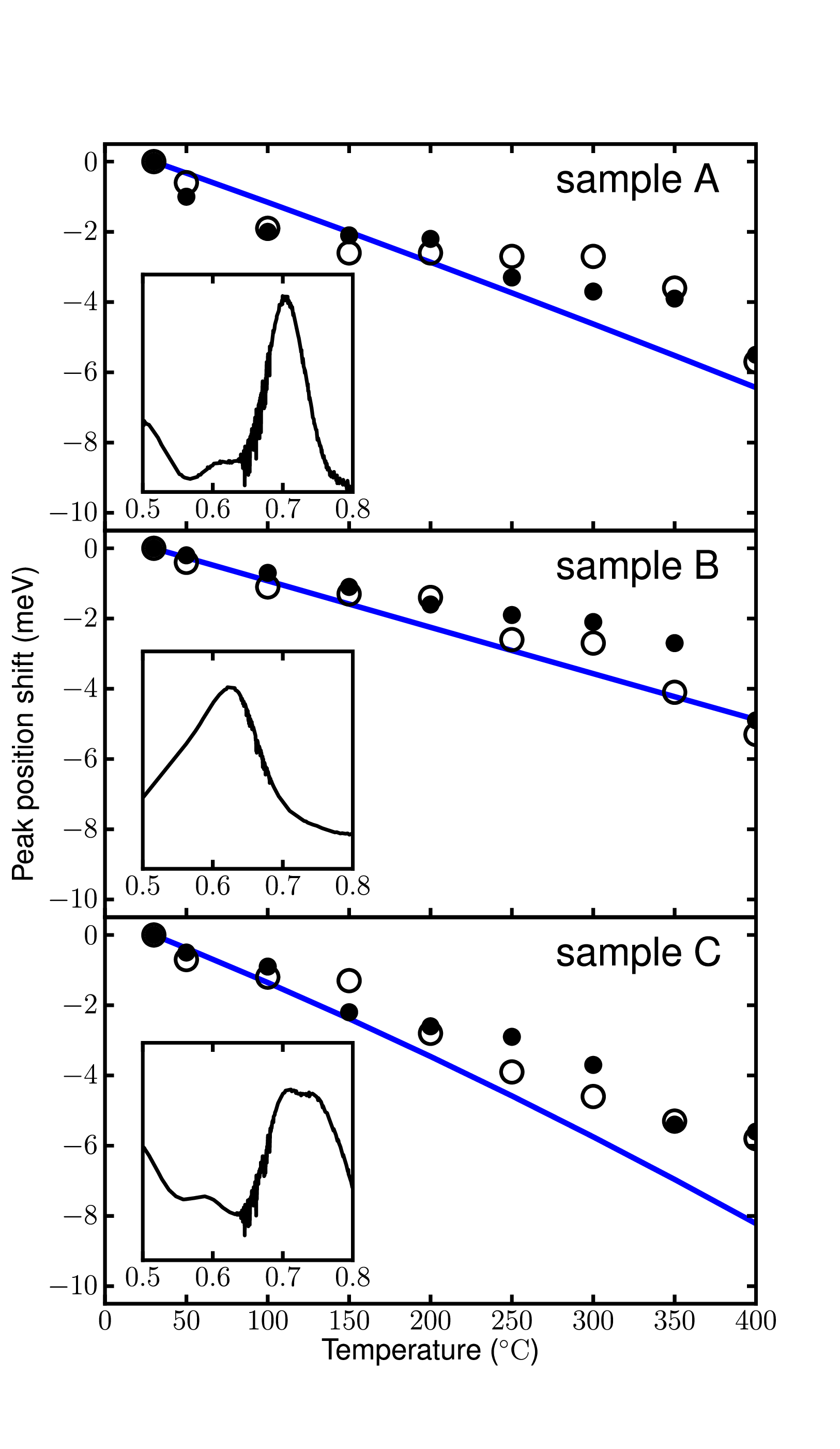}
 \caption{Temperature-dependent shift in absorption peak-position for samples A to C. The peak positions are given relative to the room temperature result. The experimental data obtained by increasing the sample temperature  are given by the filled dots, while the open circles give the data when decreasing the sample temperature. The full line gives the calculated shift in peak position. The inset shows the absorption spectrum at room temperature, where the $x$-axis gives the photon energy in eV.}
\label{fig:TempAbs}
\end{figure}

Figure \ref{fig:TempAbs} shows the shift of the absorption energy for the three samples at different temperatures. To emphasize the temperature trend, the data-series are shifted to zero at room temperature. The results for sample A to C are ordered from top to bottom in the panels of the figure. The filled dots (circles) give the increasing (decreasing) temperature scan. The good coincidence between them indicate the low experimental uncertainities. The experimental peak positions shift by $\sim$6 meV as temperature is increased from room temperature to 400$^\circ$.  The full line gives the calculated shift. The insets show the measured room-temperature absorption spectra for the samples.  Spectra A and B only show a rigid shift with temperature, while spectrum C are slightly broadened.  

The calculated and experimental results are in good agreement. There is a somewhat poorer correspondence for sample C. This discrepancy could be related to presence of a double peak. We attribute multiple peaks to mono-layer fluctuations. A well that is one mono-layer thinner (wider) shifts the calculated peak position by 40 (30) meV. This shift explains the shape of the spectra in the inset of Fig. \ref{fig:TempAbs}.

The temperature-dependent shift in transition energy is most affected by the shift in band offset, but the strain-induced piezoelectric effect is also essential. By only including the temperature dependent band offset, we get 60, 67, and 80 \% respectively of the full energetic shift for sample A to C. The higher percentage for sample C stem from a larger penetration into the left AlN barrier, and therefore an increased sensitivity to the band offset. The remaining part of the shift come from the piezoelectric effect. The energetic shift caused by the redistribution of the electrons is negligible, because of the large intersubband transition energies.

The documented temperature insensitivity for intersubband transition energies in AlN/GaN MQWs is very promising for intersubband devices made with these materials. 

The authors are grateful to F. Capasso, Q. J. Wang and C. Pl\"ugl for helpful discussions.  X. Liu is acknowledged for assistance in growth and S. Lourdudoss for supplying substrates.  Financial support was recieved from Vinnova through the grant distribution "Banbrytande IKT 2007".


\end{document}